\documentclass[sigconf]{acmart}
\usepackage{booktabs}
\usepackage{array}
\usepackage{enumitem}
\usepackage{pbalance}

\AtBeginDocument{%
  }

\setcopyright{cc}
\copyrightyear{2026}
\acmYear{2027}
\acmDOI{XXXXXXX.XXXXXXX}
\acmConference[SIGCSE TS 2027]{The Technical Symposium on Computer Science Education}{February 17--20,
  2027}{Sacramento, CA}
\acmISBN{978-1-4503-XXXX-X/2018/06}

\begin{document}
\raggedbottom

\title{When LLM Tutoring Responses Work: Evidence from Student Programming Conversations}





\author{Mohammad Fahim Abrar}
\email{fahim@udel.edu}
\affiliation{%
  \institution{University of Delaware}
  \city{Newark}
  \state{Delaware}
  \country{USA}
}

\author{Shyala Sharmin}
\email{shayla@udel.edu}
\affiliation{%
  \institution{University of Delaware}
  \city{Newark}
  \state{Delaware}
  \country{USA}
}

\author{Roghayeh Leila Barmaki}
\email{rlb@udel.edu}
\affiliation{%
  \institution{University of Delaware}
  \city{Newark}
  \state{Delaware}
  \country{USA}
}

\renewcommand{\shortauthors}{Abrar et al.}

\begin{abstract}

As students increasingly use LLM tutors in computer science education, one question becomes especially important: what kind of response helps a student continue productively? Prior work has studied how students use LLMs in computer science education, but less is known about how tutoring response styles are associated with student follow-up across programming help-seeking contexts. This paper analyzes StudyChat (UMass, 2026), a public dataset of student and ChatGPT tutoring conversations from an artificial intelligence course. We transformed StudyChat into 16,851 assistant-response interactions from 203 students and 2,214 conversations. Using local LLM-assisted annotation with Gemma 4, we labeled student help-seeking situations, student state, assistant response style, and student next-turn outcome. Human validation showed 82\% agreement with the LLM-assisted labels (Cohen's $\kappa=.74$). We analyzed productive continuation and unresolved continuation across the full dataset and across help-seeking contexts. Globally, response style was significantly associated with productive continuation, $\chi^2(7)=100.39$, $p<.001$, $V=.078$, and unresolved continuation, $\chi^2(7)=125.77$, $p<.001$, $V=.087$, though effect sizes were small. Verification feedback had the highest productive-continuation rate (82.4\%), while direct answers had the lowest (62.7\%). Descriptively, response-style score ranges were smallest in low-confusion conceptual contexts (.017) and largest in high-cognitive-load contexts (.203). More detailed comparisons showed situation-dependent response patterns. For example, stepwise guidance was followed by greater confusion decrease in high-cognitive-load code requests, while direct answers were followed by more unresolved continuation in high-load debugging. These findings support context-aware evaluation and design of AI tutoring responses for programming education.
\end{abstract}



\begin{teaserfigure}
 \includegraphics[width=\textwidth]{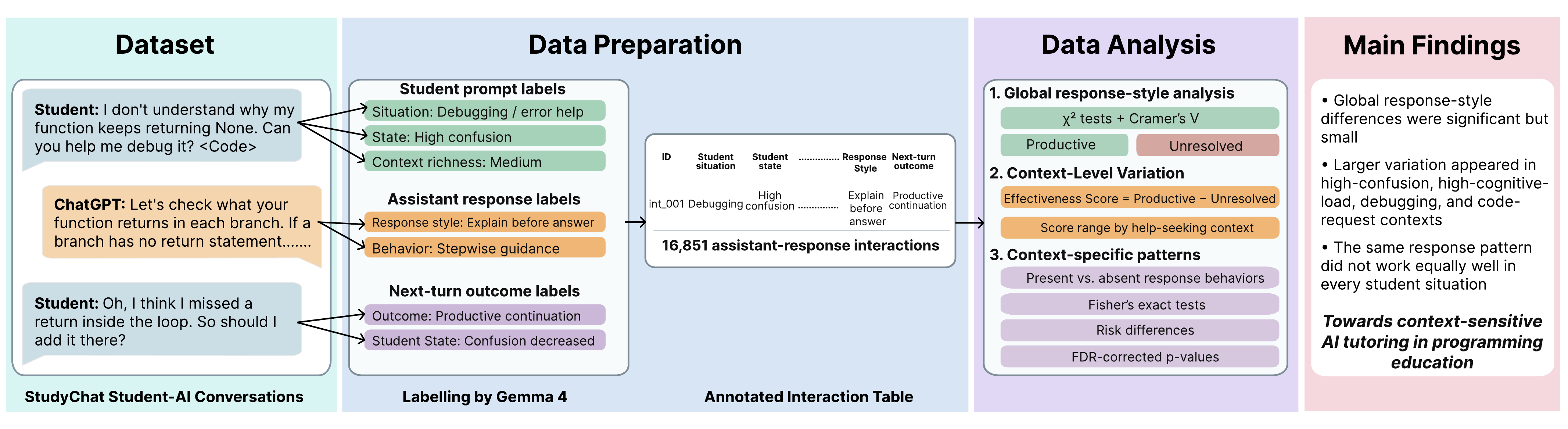}
  \Description{Flow diagram showing StudyChat conversations transformed into assistant-response interactions, annotated for help-seeking situation, student state, response style, and next-turn outcome, then analyzed for response patterns.}
  \caption{Overview of the analysis pipeline. Authentic student-ChatGPT tutoring conversations from StudyChat were transformed into assistant-response interactions, annotated with Gemma 4 for student help-seeking situation, student state, assistant response style, and next-turn outcome, and analyzed to identify response patterns in programming help-seeking.}
  \label{fig:teaser}
\end{teaserfigure}



\maketitle

\section{Introduction}

Computer Science students increasingly turn to large language models (LLMs) when they are stuck, confused, or trying to move forward in programming tasks. In these moments, the response style of the assistant matters: a student may receive a direct answer, an explanation before the answer, a scaffolded hint, stepwise guidance, generated code, verification feedback, or debugging diagnosis. Yet it is not clear whether any one response style is generally more effective, or whether its usefulness depends on what the student is trying to do and what state the student is in.

This problem is especially important in programming education because \textit{student help-seeking situations} vary widely \cite{sheese2024patterns,kurniawan2021steps}. A student asking for code generation may need a different kind of response than a student debugging an error, checking an answer, or trying to understand a concept. Similarly, a student showing high confusion or high cognitive load may respond differently to the same assistant response style than a student asking a low confusion conceptual question. Treating all student-AI interactions as equivalent can therefore lead to poorly calibrated AI tutoring responses, such as providing brief or direct completions when students need diagnostic guidance, structure, or scaffolding. Understanding these context-specific patterns is a necessary step toward adaptive LLM tutoring systems that respond to help-seeking needs in more learning-centered ways.

Prior work has examined LLMs in programming education from several perspectives, including their ability to solve programming tasks, respond to novice programmers' help requests, and generate next-step hints \cite{hellas2023help, kiesler2023intro, roest2023hints}. These studies show both the promise and limitations of LLM-based programming assistance. However, less is known about how different LLM response styles are associated with students' help-seeking situations across naturally occurring student-AI programming dialogues.

To investigate this question, we use the StudyChat dataset, a public dataset of real student dialogues with ChatGPT in a university-level artificial intelligence course \cite{McNichols2026StudyChat}. Across the Fall 2024 and Spring 2025 semesters, 203 of 295 enrolled students consented to data collection, producing 2,214 student and LLM conversations across seven programming assignments and 16,851 utterances. StudyChat is appropriate for this analysis because it is a naturalistic dataset that captures longitudinal, course-embedded interactions as students worked on programming assignments, rather than isolated benchmark prompts. This structure allows us to examine not only what kind of response the assistant provided, but also how students continued in the next turn after receiving different kinds of responses. This paper addresses the following research questions:



\begin{description}[leftmargin=0pt, labelindent=0pt, labelsep=0.5em]
    \item[\textbf{RQ1:}] How are LLM tutoring response styles associated with students' productive and unresolved next-turn engagement in student-AI programming dialogues?

    \item[\textbf{RQ2:}] How does LLM response effectiveness vary across programming help-seeking contexts, including conceptual understanding, code generation, debugging, confusion, and high cognitive load?
\end{description}


\noindent We make the following contributions:

\begin{itemize}[leftmargin=*, itemsep=0.2em]
    \item We analyze LLM tutoring response styles in natural student-AI programming dialogues from StudyChat, covering 16,851 assistant-response interactions.

    \item We show that global response-style differences in next-turn engagement are statistically significant but small, which is consistent with the naturalistic, student-led nature of StudyChat.

    \item We identify response-style patterns across programming help-seeking contexts.
\end{itemize}

\section{Related Work}

\subsection*{LLM Tutors in Computer Science Education}

Computer science education research has moved from asking whether code-generation models can solve programming tasks to examining how LLMs can support student learning. Denny et al. showed that students can use natural-language prompting and iterative revision with GitHub Copilot to solve many CS1 problems ~\cite{Denny2023Copilot}. More recent systems, including CS50.ai, customized ChatGPT tutors, Iris, CodeAid, and 61A Bot, incorporate course context, guardrails, hints, and tutor-like responses to support learning rather than simply provide solutions \cite{Liu2024CS50,Frazier2024Customizing,Bassner2024Iris,Kazemitabaar2024CodeAid,ZamfirescuPereira2025Bot}. LLM use also extends beyond CS1. Osorio et al. studied AI use in a database course ~\cite{Osorio2025Database}. Shah et al. examined Copilot use with large code bases ~\cite{Shah2025CopilotLarge}. Alpizar-Chacon and Keuning investigated AI use in advanced web development ~\cite{AlpizarChacon2025WebDev}. While this work shows broad adoption across programming contexts, less is known about how specific assistant response styles are followed by students' immediate next state.

\subsection*{Pedagogical Response Strategies}

Research on scaffolding, help-seeking, and intelligent tutoring provides a foundation for studying LLM response style. Wood et al.~\cite{Wood1976Tutoring} described tutoring as scaffolding that helps learners continue problem solving while maintaining task ownership. Aleven et al. showed that help can support learning, but its effectiveness depends on timing, interpretation, and student use ~\cite{Aleven2016Help}. In computer science education, help-seeking is shaped by task context, available support, social concerns, and perceived usefulness. Price et al. examined programming help-seeking with human and computer tutors, while Cohen et al. showed that social and classroom contexts influence help-seeking~\cite{Price2017HelpSeeking, Cohen2024SocialHelp}. In the generative AI era, Penney et al. found that students may view AI chatbots as accessible and less intimidating, while Denny et al. showed that students value timely AI support that preserves learning opportunities ~\cite{Penney2025HelpPref, Denny2024Characteristics}. Building on this work, we examine whether response styles are followed by different student states.

\subsection*{Analyzing Student-LLM Conversations}

Conversation-level datasets enable analysis of LLM tutoring as interaction rather than isolated prompts. McNichols et al. introduced StudyChat, which captures real student interactions with a ChatGPT-based tutoring system in an artificial intelligence course ~\cite{McNichols2026StudyChat}. Related dialogue-analysis work, such as MathDial shows that strong answer generation does not necessarily imply strong tutoring, because a response may solve the problem while revealing too much or failing to guide reasoning \cite{Macina2023MathDial}. Prior work has also studied debugging interactions, guardrails, escalation mechanisms, perceptions, and usage patterns. Yang et al. examined novice help-seeking during debugging with an AI tutor ~\cite{Yang2024Debugging}. Phung et al. studied an instructor-in-the-loop system for cases where AI assistance may be insufficient ~\cite{Phung2026Closing}. However, prior work generally does not systematically connect a specific assistant response to the student's next-turn state across different student help-seeking situations. Our work addresses this gap by analyzing StudyChat student-AI dialogues.

\section{Methods}
\subsection*{Dataset}

We used the StudyChat dataset, a real-world collection of student interactions with an LLM-powered tutoring chatbot in an upper-division university artificial intelligence course at the University of Massachusetts Amherst~\cite{McNichols2026StudyChat}. The dataset was collected longitudinally across Fall 2024 and Spring 2025 while students worked over 15 weeks on seven Python-based programming assignments. Students had completed foundational computer science course sequences before taking the course. The original dataset includes interactions from 203 consenting students across 2,214 student and ChatGPT conversations. These conversations capture authentic student help-seeking situations. For this study, we transformed the raw conversations into an assistant-response-level interaction dataset. Each row contained the preceding student prompt, the assistant response, prior conversational context when available, the next student prompt when available, and metadata such as \texttt{userId}, \texttt{chatId}, semester, assignment topic, turn index, total turns, and turn position. The preprocessing pipeline constructed 16,851 interaction rows from 203 unique students.

\subsection*{Data Preparation}

To analyze the tutoring conversations at scale, we annotated each AI assistant-response interaction using a local LLM, \texttt{Gemma4} with 26B parameters. The annotation was run on a workstation with an NVIDIA RTX 3090 GPU with 24 GB VRAM, 64 GB system RAM, and an Intel Core i9 3.70 GHz processor. Each interaction was labeled using three structured prompts. The \textit{first prompt }classified the student's situation before the AI assistant's response. The \textit{second prompt} classified the assistant's response style. The \textit{third prompt} classified what happened in the student's following turn. The prompts required the model to return structured JSON only and avoid unsupported guesses. Two authors independently labeled 100 randomly sampled consecutive student and LLM dialogues. There was an 82\% agreement with the LLM-assisted labels, and Cohen’s kappa was 0.74.

The student help-seeking situation annotation captured what kind of help the student was seeking. We used labels for common programming help-seeking situations, including conceptual questions, debugging or error help, code generation or completion, verification or checking, clarification requests, assignment interpretation, repeated confusion, direct-answer seeking, and other less frequent requests. Conceptual questions were identified when students asked for explanations or conceptual understanding. Debugging or error help was identified when students reported errors, bugs, exceptions, or failed outputs. Code generation or completion was identified when students asked the tutor to write, complete, modify, or generate code. Verification or checking was identified when students asked whether an answer, code fragment, or approach was correct. Clarification requests were identified when students asked the tutor to rephrase, simplify, or explain a prior response. Assignment interpretation was identified when students asked how to understand assignment requirements. For the main response-effectiveness analysis, we focused on the six most common and analytically meaningful situations: code generation or completion, conceptual question, debugging or error help, clarification request, verification or checking, and assignment interpretation. Less frequent categories were retained during annotation but treated as other categories in the analysis.

The assistant-response annotation captured how the LLM responded pedagogically. We labeled response styles as direct answer, explanation before answer, scaffolded hint, stepwise guidance, debugging diagnosis, code generation, verification feedback, or other. Direct-answer responses gave the answer or solution with little explanation. Explanation-before-answer responses explained the reasoning or concept before giving an answer. Scaffolded hints provided partial guidance or prompts without immediately giving the full solution. Stepwise guidance organized help as ordered steps or procedures. Debugging diagnosis identified a likely bug, error source, or failure mechanism. Code-generation responses primarily provided code or code-like solution content. Verification-feedback responses checked the student's work, confirmed correctness, or gave corrective feedback. These labels allowed us to compare whether different response styles were followed by productive continuation or unresolved continuation.

In addition to the primary situation and response-style labels, the annotation prompts produced secondary variables used to construct the analysis measures. Student state included confusion level, cognitive-load signal, frustration signal, context richness, and answer-seeking level. AI assistant response included pedagogical strategy, response form, direct solution giving, scaffolding, example use, and stepwise guidance. Next-turn variables captured whether the student's following message showed productive continuation, repeated the same issue, indicated a change in confusion, or suggested that the conversation was resolved. These variables were used to define the help-seeking contexts, response style comparisons, and next-turn outcome measures reported in the analysis.

\subsection*{Data Analysis}

All analyses were conducted at the assistant-response level. Each row represented one assistant response, its annotated response properties, the student's help-seeking context, and the student's immediate following turn. Interactions without an available following student turn were not used in next-turn outcome-rate calculations. Because the data are observational, we report associations between assistant responses and the following student outcomes rather than causal effects.

\subsubsection*{Analysis Variables and Outcome Construction}

We constructed the analysis table by merging student-situation labels, student-state labels, assistant-response labels, and next-turn outcome labels for each interaction. Student help-seeking contexts combined the student's task situation with state information such as confusion, cognitive load, context richness, and frustration. These contexts were used to examine response patterns across different student needs rather than only across broad task categories.

Assistant response styles were grouped into the main categories used in the results: explanation before answer, scaffolded hint, code generation, stepwise guidance, direct answer, verification feedback, and debugging diagnosis. Less frequent response-style labels were collapsed into a sparse category and were not used for context-level comparisons unless the cell size was sufficient.

The main outcomes were determined from the student's immediate next turn after each AI assistant response. \textit{Productive continuation} was coded when the following student message showed productive engagement, such as continuing the task, asking a meaningful follow-up question, applying the response, requesting verification, or indicating progress. \textit{Unresolved continuation} was coded when the following student message showed that the issue remained unresolved, such as repeated confusion or a repeated request about the same problem.

For context-level analyses, we also used additional student-state data. \textit{Cognitive load} captured whether the student prompt showed signs of task complexity or overload. \textit{Confusion decreased} was coded when the following student turn indicated reduced confusion or clearer understanding after the AI assistant response. \textit{Same issue repeated} was coded when the student repeated the same problem or request in the following turn. These variables allowed the analyses to compare response styles and assistant response styles across student states, using both favorable and unfavorable student next-turn outcomes.

\subsubsection*{Global Response-Style Associations}

To examine global response-style patterns, we tested whether assistant response style was associated with productive continuation and unresolved continuation across the full dataset. We used chi-square tests of independence and reported the chi-square statistic, $p$-value, and Cramer's $V$. Outcome rates for each response style were then plotted to show the productive-continuation rate and unresolved-continuation rate for each style.

\subsubsection*{Response-Style Variation}

To examine whether response-style differences varied across student help-seeking contexts, we calculated an effectiveness score for each response style within each context. Let \(PCR\) denote the productive-continuation rate and \(UCR\) denote the unresolved-continuation rate. The effectiveness score was calculated as
\[
ES = PCR - UCR .
\]

This score summarizes the balance between a favorable following outcome and an unfavorable next-turn outcome. For each help-seeking context, the score range was calculated as
\[
SR = \max(ES) - \min(ES).
\]

Only response-style cells with at least 50 observations were included in the score-range calculation. This threshold was used to avoid allowing very small cells to determine the highest or lowest observed score within a context. The resulting score ranges were used to identify contexts where response-style outcome differences were relatively small or relatively large.

\subsubsection*{Assistant Response Style Comparisons}

We also examined specific assistant response style within the student help-seeking situation. For each student situation, AI response style, and outcome combination, responses were divided into two groups: \textit{present} and \textit{absent}. \textit{Present} means that the annotated assistant response style was detected in the response. \textit{Absent} means that the same response style was not detected in the response. For each group, we calculated the outcome rate in the student's following turn.

The difference column reports the risk difference:
\[
RD = R_{\mathrm{present}} - R_{\mathrm{absent}},
\]
where \(R_{\mathrm{present}}\) and \(R_{\mathrm{absent}}\) are the outcome rates when the response style was present or absent.

A positive difference means that the outcome rate was higher when the response style was present. A negative difference means that the outcome rate was lower when the response style was present. The interpretation depends on the outcome: higher values are favorable for productive continuation, and confusion decreased, but unfavorable for unresolved continuation and same-issue repetition.

For each comparison, we used Fisher's exact test to compare the outcome distribution between the present and absent groups. Because many comparisons were tested, we adjusted the resulting $p$-values using a false-discovery-rate (FDR) correction. The adjusted value is reported as \(p_{\mathrm{FDR}}\). This correction reduces the chance of treating a comparison as significant simply because many related comparisons were examined. The results table reports selected comparisons that passed FDR correction and had sufficient observations for both the present and absent groups.

\section{Results}

\subsection*{Global Response-Style Patterns}
\begin{figure}[t]
    \centering
    \includegraphics[width=\columnwidth]{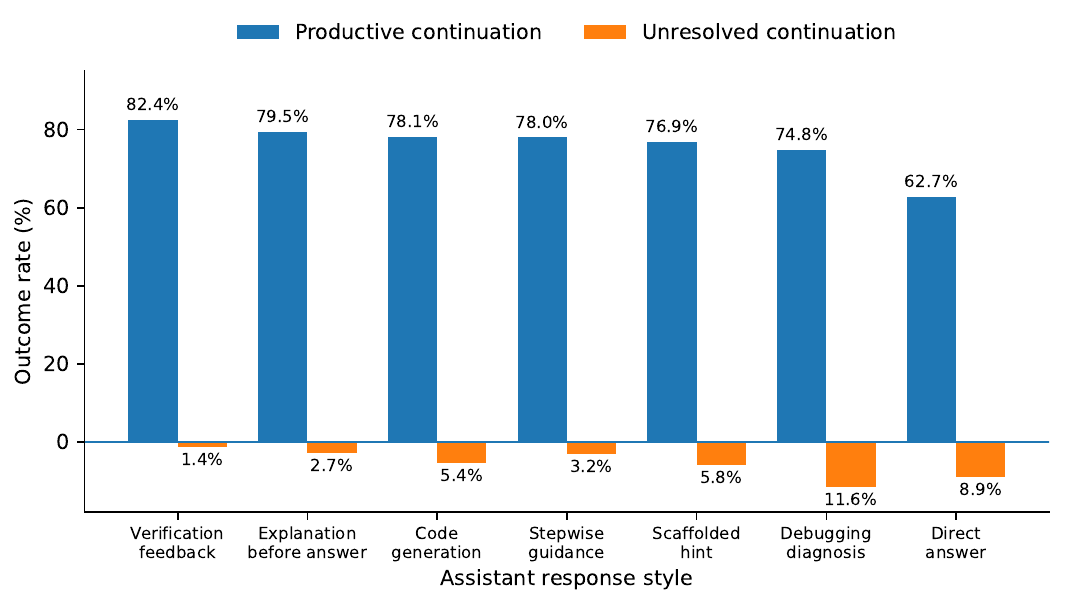}
    \Description{Bar chart comparing productive continuation and unresolved continuation rates for each assistant response style. Productive continuation is shown above zero and unresolved continuation below zero.}
    \caption{Productive continuation and unresolved continuation rates by assistant response style. Productive continuation is plotted above zero, while unresolved continuation is plotted below zero to indicate that lower unresolved continuation is more favorable.}
    \label{fig:global-response-outcomes}
\end{figure}

Response style was significantly associated with productive continuation, $\chi^2(7)=100.39$, $p<.001$, $V=.078$, and with unresolved continuation, $\chi^2(7)=125.77$, $p<.001$, $V=.087$. The Cramer's $V$ values indicate small effect sizes. Figure~\ref{fig:global-response-outcomes} shows productive continuation above the zero line and unresolved continuation below the zero line. Verification feedback had the highest productive-continuation rate at 82.4\%, followed by explanation before answer at 79.5\%, code generation at 78.1\%, stepwise guidance at 78.0\%, scaffolded hint at 76.9\%, and debugging diagnosis at 74.8\%. Direct-answer responses had the lowest productive-continuation rate at 62.7\%. Unresolved continuation rates were lower overall. Verification feedback had an unresolved-continuation rate of 1.4\%, followed by explanation before answer at 2.7\%, stepwise guidance at 3.2\%, code generation at 5.4\%, scaffolded hint at 5.8\%, direct answer at 8.9\%, and debugging diagnosis at 11.6\%.

\subsection*{Response-Style Variation}

\begin{figure}[t]
    \centering
    \includegraphics[width=\columnwidth]{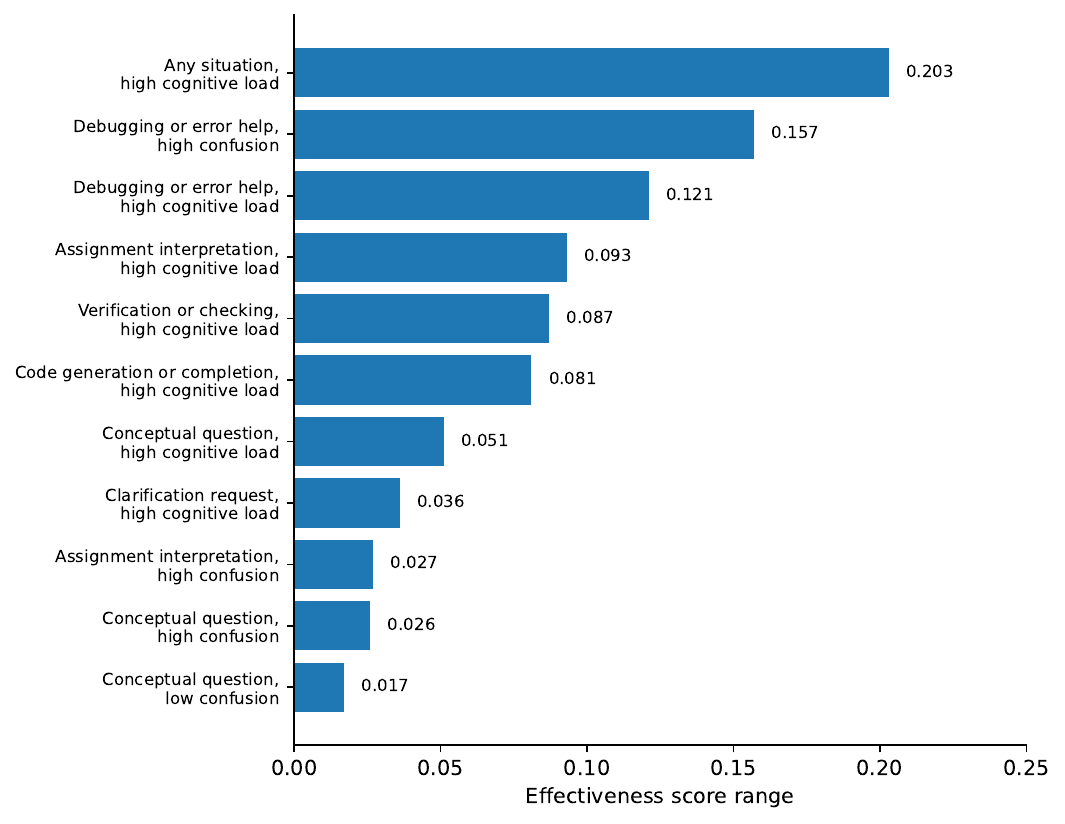}
    \Description{Bar chart showing response-style effectiveness score ranges across student help-seeking contexts, with larger ranges for higher cognitive load, debugging, and high-confusion contexts.}
    \caption{Response-style effectiveness score ranges across student help-seeking contexts. The score range is the difference between the highest and lowest observed effectiveness score among other response styles.}
    \label{fig:help-seeking-score-ranges}
\end{figure}

Figure~\ref{fig:help-seeking-score-ranges} shows the response-style effectiveness score range across student help-seeking contexts. Each context combines a student situation with a student state, and the score range represents the difference between the highest and lowest observed effectiveness scores among response styles with at least 50 observations.

The smallest score ranges appeared in conceptual contexts. Low-confusion conceptual questions had a score range of .017, and high-confusion conceptual questions had a range of .026. Conceptual questions with high cognitive load also showed a modest range of .051.

Larger score ranges appeared when student state reflected higher task or cognitive demand. The overall high-cognitive-load context had the largest range at .203. Debugging or error-help contexts also showed larger ranges, with .157 under high confusion and .121 under high cognitive load. High-cognitive-load assignment interpretation showed a moderate range of .093. These results show that response-style variation was not uniform across student help-seeking contexts. 

\subsection*{AI Response-Style Patterns}

Table~\ref{tab:key-triad-patterns} presents selected significant AI response-style patterns within student help-seeking contexts. In high-confusion prompts across any student situation, direct-answer responses had a same-issue-repeated rate of 10.3\%, compared with 5.3\% when direct-answer responses were absent. In high-cognitive-load prompts across any student situation, scaffolded-hint responses had a confusion-decreased rate of 28.6\%, compared with 19.8\% when scaffolded-hint responses were absent. In high-cognitive-load code requests, stepwise guidance had a confusion-decreased rate of 24.8\%, compared with 15.6\% when stepwise guidance was absent. In low-confusion code requests, stepwise guidance had a productive-continuation rate of 82.3\%, compared with 76.6\% when stepwise guidance was absent. In low-confusion conceptual questions, scaffolded-hint responses had a confusion-decreased rate of 24.9\%, compared with 16.7\% when scaffolded-hint responses were absent.  In high-confusion debugging contexts, direct-answer responses had an unresolved-continuation rate of 15.0\%, compared with 8.3\% when direct-answer responses were absent. In high-cognitive-load debugging contexts, direct-answer responses had an unresolved-continuation rate of 15.4\%, compared with 8.3\% when direct-answer responses were absent.

\begin{table*}[t]
\centering
\caption{Selected significant AI response-style patterns within student help-seeking contexts. Present and absent indicate outcome rates when the response style was or was not detected in the assistant response. All rows passed FDR correction.}
\label{tab:key-triad-patterns}
\footnotesize
\renewcommand{\arraystretch}{1.12}
\begin{tabular}{lllrrrrrrr}
\toprule
Student Situation & Student State & AI Response Style & Student Next-Turn Outcome
& $n_{\mathrm{present}}$ & $n_{\mathrm{absent}}$
& Present & Absent & Diff. & $p_{\mathrm{FDR}}$ \\
\midrule

Any situation & High confusion & Direct answer & Same issue repeated
& 881 & 2225 & 10.3\% & 5.3\% & +5.1 & $<.001$ \\

Any situation & High cognitive load & Scaffolded hint & Confusion decreased
& 3841 & 1331 & 28.6\% & 19.8\% & +8.9 & $<.001$ \\

Code generation or completion & High cognitive load & Stepwise guidance & Confusion decreased
& 836 & 410 & 24.8\% & 15.6\% & +9.2 & .001 \\

Code generation or completion & Low confusion & Stepwise guidance & Productive continuation
& 2597 & 2016 & 82.3\% & 76.6\% & +5.7 & $<.001$ \\

Conceptual question & Low confusion & Scaffolded hint & Confusion decreased
& 1020 & 1317 & 24.9\% & 16.7\% & +8.2 & $<.001$ \\

Debugging or error help & High confusion & Direct answer & Unresolved continuation
& 486 & 1108 & 15.0\% & 8.3\% & +6.7 & .001 \\

Debugging or error help & High cognitive load & Direct answer & Unresolved continuation
& 520 & 1200 & 15.4\% & 8.3\% & +7.1 & $<.001$ \\

\bottomrule
\end{tabular}
\end{table*}

\section{Discussion}
\subsection*{RQ1: Global Response-Style Patterns}

The global results show that assistant response style was associated with students' next-turn state, but the small Cramer's $V$ values indicate that these associations were limited in size. This suggests that response style has measurable relevance, but global response-style differences should not be interpreted as a universal ranking of tutoring quality. Several response styles had relatively high productive-continuation rates, including verification feedback, explanation before answer, code generation, and stepwise guidance. At the same time, unresolved-continuation rates were generally low across styles, which indicates that many assistant responses were followed by some form of continued engagement rather than immediate repetition or breakdown.

The relatively favorable global pattern for verification feedback and explanation-oriented responses is consistent with prior work emphasizing the role of feedback, explanation, and guided assistance in learning environments \cite{hattie2007feedback,shute2008feedback}. In programming-specific LLM research, prior studies have also shown that LLMs can provide useful explanations and feedback for students' programming help requests, while still producing responses that may be incomplete, misleading, or overly solution-focused \cite{hellas2023help,kiesler2023formative}. Our results align with this mixed view: some response styles were followed by more favorable next-turn outcomes, but the differences were not large enough to justify treating one style as generally best.

Direct-answer responses had the lowest productive-continuation rate, while debugging diagnosis had the highest unresolved-continuation rate. These patterns should be interpreted cautiously. A direct answer may sometimes close a simple request, but it may also reduce opportunities for students to articulate reasoning or continue productively. Similarly, debugging diagnosis may appear in more difficult interactions where students are already facing errors or uncertainty. Therefore, the global results provide a useful baseline, but they also show the limitation of evaluating LLM tutoring responses without considering the student's help-seeking context.

\subsection*{RQ2: Response Effectiveness Across Help-Seeking Contexts}

The context-level results show that response-style effectiveness varied more strongly in some help-seeking contexts than in others. The smallest score ranges appeared in conceptual contexts, including low-confusion conceptual questions and high-confusion conceptual questions. Even conceptual questions with high cognitive load showed only a modest range. This may indicate that, for many conceptual questions, several response styles can still provide enough information for the student to continue, as long as the response addresses the central idea.

Larger score ranges appeared when the student state reflected higher cognitive demand, especially high cognitive load and debugging-related contexts. This pattern is consistent with cognitive load theory, which emphasizes that learners have limited working-memory resources when processing complex tasks \cite{sweller1988cognitive,paas2003cognitive}. Programming tasks can already impose substantial cognitive demand because students must coordinate syntax, semantics, program state, and problem goals. When students are also confused or overloaded, the same response style may become more consequential because the response can either reduce the burden of organizing the task or leave the student with the same unresolved difficulty.

The high-confusion result is also important because confusion is not simply a negative state. Prior work on learning and affect suggests that confusion can be productive when it is resolved, but may become unproductive when it persists without sufficient resolution \cite{dmello2014confusion}. In our results, high-confusion contexts had the largest response-style score range, and direct-answer responses were followed by higher same-issue repetition. This suggests that when students are highly confused, simply providing the answer may not address the source of confusion that shaped the original request.

The debugging and code-request patterns point in a similar direction. Debugging requires students to reason about program behavior, locate errors, and connect symptoms to causes. Prior work in programming education has noted that novice programmers often need structured problem-solving processes to manage these demands \cite{kurniawan2021steps}. In our results, stepwise-guidance responses in high-cognitive-load code requests were followed by greater confusion decrease, while direct-answer responses in high-confusion and high-cognitive-load debugging contexts were followed by higher unresolved continuation. These patterns suggest that, in complex programming situations, the useful part of a response may not be only whether it gives an answer, but whether it helps the student organize the problem, reason through the issue, and move past the immediate impasse.

Overall, RQ2 shows that response-style effectiveness should be interpreted relative to the student's help-seeking context. The global results showed only small overall differences, but the context-level results revealed larger variation when students appeared confused, cognitively loaded, or engaged in complex programming tasks. This supports a context-sensitive interpretation of LLM tutoring responses: the same response pattern may be followed by different outcomes depending on the cognitive, affective, and task demands present in the student's context.

\subsection*{Limitations and Future Directions}

Like any preliminary study, this study has several limitations. First, StudyChat is a naturalistic, student-led dataset, so help-seeking contexts and response styles were not evenly distributed or experimentally assigned. This may reflect authentic course use, but it also means that some patterns may be shaped by dataset characteristics such as assignment design and chatbot usage norms. Future studies should examine whether similar patterns appear across different courses, institutions, and tutoring platforms. Second, next-turn outcomes may not fully capture response effectiveness because students may have resolved the issue on their own using other resources beyond the LLM, moved to another task, or disengaged for reasons not visible in the dialogue. Future work should therefore examine longer-term outcomes such as learning gains, debugging success, code improvement, and student confidence. Finally, the analyses are observational, so the results show associations between assistant response patterns and students' next-turn state rather than causal effects. Controlled studies are needed where response style can be experimentally varied within the same student situation. Building on these findings, our longer-term aim is to develop adaptive tutoring systems that detect student help-seeking situations and state in real time and select response strategies accordingly, moving beyond a single global tutoring style toward context-sensitive support for programming help.

\section{Conclusion}
In this paper, we analyzed how LLM tutoring response styles are associated with students' next-turn engagement in authentic student-AI programming dialogues. Global response-style differences were statistically significant but small, while larger differences appeared across help-seeking contexts involving confusion, cognitive load, debugging, and code requests. These findings suggest that evaluating LLM tutors requires looking beyond whether a response is correct to whether it fits the student's situation. The next challenge is not only building AI tutors that answer programming questions, but designing AI tutors that know what kind of help a student needs in the moment.

\bibliographystyle{ACM-Reference-Format}
\bibliography{main_paper}

\end{document}